\documentclass[structabstract]{aa}  
\usepackage{graphicx}
\usepackage{hyperref}
\usepackage{natbib}
\bibpunct{(}{)}{;}{a}{}{,}
\usepackage{txfonts}

\newcommand\FeI{\mbox{Fe\,\textsc{i}}} 
\newcommand\NaID{\mbox{Na\,\textsc{i}\,D\ensuremath{_1}}} 

\begin{document}
\title{Observations of solar scattering polarization\\
at high spatial resolution}


\author{F. Snik\inst{1}
\and A. G. de Wijn\inst{2}
\and K. Ichimoto\inst{3}
\and C. E. Fischer\inst{1}
\and C. U. Keller\inst{1}
\and B. W. Lites\inst{2}
}

\institute{Sterrekundig Instituut Utrecht, Princetonplein 5, 3584 CC, Utrecht, the Netherlands\\
\email{f.snik@astro.uu.nl}
\and High Altitude Observatory, National Center for Atmospheric Research, P.O. Box 3000, Boulder, CO, 80307-3000, U.S.A.
\and Kwasan and Hida Observatories, Kyoto University, Kurabashira Kamitakara-cho, Takayama-city, Gifu, 506-1314, Japan}



\abstract
{The weak, turbulent magnetic fields that supposedly permeate most of the solar photosphere are difficult to observe, because the Zeeman effect is virtually blind to them.
The Hanle effect, acting on the scattering polarization in suitable lines, can in principle be used as a diagnostic for these fields.
However, the prediction that the majority of the weak, turbulent field resides in intergranular lanes also poses significant challenges to scattering polarization observations because high spatial resolution is usually difficult to attain.}
{We aim to measure the difference in scattering polarization between granules and intergranules.
We present the respective center-to-limb variations, which may serve as input for future models.}
{We perform full Stokes filter polarimetry at different solar limb positions with the CN band filter of the \textit{Hinode}-SOT Broadband Filter Imager, which represents the first scattering polarization observations with sufficient spatial resolution to discern the granulation.
\textit{Hinode}-SOT offers unprecedented spatial resolution in combination with high polarimetric sensitivity.
The CN band is known to have a significant scattering polarization signal, and is sensitive to the Hanle effect. 
We extend the instrumental polarization calibration routine to the observing wavelength, and correct for various systematic effects.}
{The scattering polarization for granules (i.e.,~regions brighter than the median intensity of non-magnetic pixels) is significantly larger than for intergranules.
We derive that the intergranules (i.e.,~the remaining non-magnetic pixels) exhibit $(9.8\pm3.0)\%$ less scattering polarization for $0.2<\mu\le0.3$, although systematic effects cannot be completely excluded.}
{These observations constrain MHD models in combination with (polarized) radiative transfer in terms of CN band line formation, radiation anisotropy, and magnetic fields.}


\authorrunning{Snik et al.}
\titlerunning{Solar scattering polarization at high spatial resolution.}

\maketitle
%

\section{Introduction}
With the increasing sensitivity of solar spectropolarimeters, a substantial amount of knowledge about quiet Sun magnetic fields is obtained using the Zeeman effect in suitable spectral lines (for a recent overview, see \citealp{dewijnreview}).
However, the Zeeman effect is virtually blind to turbulent (i.e., mixed-polarity) magnetic fields because the polarization signals in such observations cancel out for opposite polarities within a resolution element.
A different line polarization effect can be used to probe these weak, turbulent magnetic fields: the Hanle effect, which acts on lines with scattering polarization (see, e.g., \citealp{TBreview}).
This scattering polarization is observed near the solar limb, where the scattering angle is close to 90$^\circ$.
However, the degree of polarization (Stokes $Q/I$, with $+Q$ parallel to the local limb) is low: $\sim$$0.01$ down to $10^{-5}$ \citep{sss}.
In general, the Hanle effect constitutes a depolarization of the line by weak magnetic fields of $\sim$$0$--$100~\mathrm{G}$.

\begin{figure}[t]
	\includegraphics[]{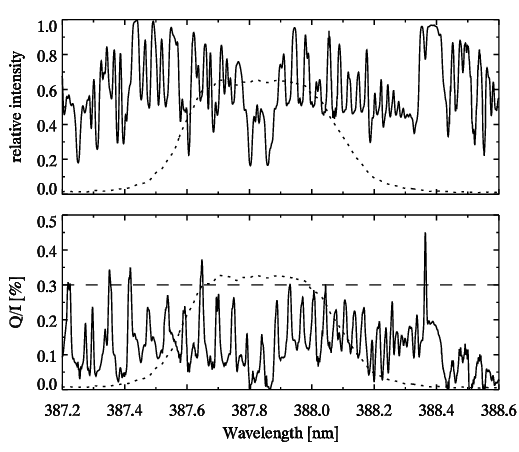}
	\caption{The intensity and scattering polarization spectrum at $\mu=0.1$ around the CN band head at 388.3 nm \citet[adapted from][]{Stenflo3883}. The position of the scattering polarization of the continuum is indicated by a dashed line and is taken from \citet{Shapiro}. Note that this polarization scale is different from the Atlas of \citet{GandorferIII}. The profile of the \textit{Hinode} CN band filter is overplotted and is displaced somewhat blueward of the bandhead. The strong lines at 387.61, 387.80 and 387.86 nm belong to \FeI. Most other lines belong to CN.}
	\label{3883}
\end{figure}

With an analysis based on the Hanle effect, \citet{TBnature} were able to infer a ubiquitous turbulent magnetic field in the quiet Sun with an average strength of $\sim$$130~\mathrm{G}$.
With an indirect method, the authors conjecture a difference in turbulent magnetic field strengths between granules and intergranules, with averages of $\sim$$15~\mathrm{G}$ and $300$--$1000~\mathrm{G}$, respectively.
This dichotomy between granules and intergranules is also observed in the MHD simulations of \citet{Voegler}, which exhibited local dynamo action by the convection.
A direct observational constraint is required to confirm these hypotheses and models, and to investigate the presence of a local dynamo in the solar photosphere.
To this end, we perform filter observations of solar scattering polarization at high spatial resolution, to resolve the granulation pattern near the limb.
High spatial resolution is very difficult to attain in current ground-based observations of scattering polarization, because adaptive optics systems do not perform well in quiet regions near the solar limb.
At the long effective exposure times of several minutes that are required to reach sufficient polarimetric sensitivity, the atmospheric seeing and solar evolution usually smear out the granulation pattern.
We present the first observations of solar scattering polarization at high spatial resolution, obtained with the Solar Optical Telescope (SOT) of the \textit{Hinode} satellite. 
The only filter position of SOT that enables these observations is the broad-band filter at the CN band head ($388~\mathrm{nm}$, see Fig.~\ref{3883}).
The CN band is usually employed to study photospheric magnetic fields through proxy-magnetometry \citep[see, e.g.,][]{UitenbroekCN}.
It is known that the continuum polarization at this blue wavelength is relatively large ($\sim$$0.3\%$ at $\mu$$\equiv$$\cos\theta$$=$$0.1$, where $\theta$ is the heliocentric angle; see \citealp{TBcont}), and that the molecular lines are sensitive to the Hanle effect \citep{Shapiro}.
Indeed, \citet{Shapiro} show that the CN band lines depolarize the continuum near the band head (their Fig.~8).
This depolarization is visible as the dips with an absolute range of $\Delta(Q/I)\approx0.15\%$ in the continuum of $Q/I$ of $\sim$$0.3\%$ in Fig.~\ref{3883}.

Our aim is to measure the center-to-limb variation (CLV) of the scattering polarization in \textit{Hinode} CN band filtergrams and to characterize the differences therein between granules and intergranules.
There are many effects that contribute to a possible difference in scattering polarization between granules and intergranular lanes:
\begin{enumerate}
	\item the chemistry of the CN molecule formation \citep[see][]{AARJTBCN} that cause formation differences of the molecular lines (see \citealp{TBnature});
	\item scattering anisotropy variations (see, e.g., \citealp{TBcont});
	\item different degrees of Hanle depolarization due to differences in (turbulent) magnetic field strength.
\end{enumerate}
With the measurements presented here it is impossible to distinguish these solar effects or estimate to what level the effects amplify or cancel one another. 
It is, however, the main goal of this research to present measurements of the scattering polarization signals in granules and intergranules in the quiet Sun to constrain future models, and to separate these signals from instrumental effects, which are potentially larger than the solar signals. 
The instrumental effects that may introduce differences in measured polarization between granules and intergranules include:
\begin{enumerate}
		\setcounter{enumi}{3}
	\item spurious signals due to solar evolution and (residual) image motion during the polarization modulation cycle;
	\item coupling of the instrumental polarization to the detector non-linearity or dark bias \cite[cf.][]{Keller96}, which results in a print-through of the intensity signal (i.e.~the granulation) in $Q/I$.
\end{enumerate}

The measured polarization signal $Q_\mathrm{meas.}$ in the quiet sun is therefore described as
\begin{equation}
	Q_\mathrm{meas.}=Q_\mathrm{real}+Q_\mathrm{i.m.}+Q_\mathrm{s.e.}+\tens{X}_{21}\,I\,(1+a\cdot I)\,,
	\label{effects}
\end{equation}
where $Q_\mathrm{i.m.}$ is the spurious signal due to image motion, and $Q_\mathrm{s.e.}$ is the spurious signal due to solar evolution.
The fourth term on the right-hand side describes the instrumental polarization, with $\tens{X}_{21}$ the $I\rightarrow Q$ component of the instrumental polarization response matrix and $a$ the normalized non-linear parameter of the detector \citep{Keller96}.

The \textit{Hinode} observations are presented in Sect.~\ref{observations}.
The data reduction and the methods to eliminate spurious polarization signals are detailed in Sect.~\ref{datareduction}.
The results are presented and discussed in Sect.~\ref{results}, and final conclusions are drawn in Sect.~\ref{conclusions}.

\section{Observations}\label{observations}

\subsection{CN band observations}
Full-Stokes observations in the CN band were obtained on Oct.~21--22, 2007 with the \textit{Hinode}-Solar Optical Telescope (SOT) \citep{Hinode1,Hinode2,Hinode3}.
These are the first scattering polarization observations with sufficient spatial resolution ($\sim$$0.2\arcsec$) to discern the granulation pattern near the limb.
Polarimetric observations with the \textit{Hinode} Broadband Filter Imager (BFI) are not offered as a standard mode, but they are possible because all the light entering the SOT instruments is polarimetrically modulated by a continuously rotating wave plate, and the light between the Narrowband Filter Imager (NFI) and the BFI is split by a polarizing beam-splitter, which therefore acts as a polarization analyzer for both instruments (only with a $90^\circ$ rotation between them).
By extending the observing software, the BFI data were polarimetrically demodulated in the same way as NFI data.

Because high spatial resolution is crucial for our analysis, we choose to observe in the CN band with small effective exposure times in order not to be very susceptible to solar evolution, and repeat the observations many times.
The modulation/demodulation mode was necessarily ``shuttered'', because the BFI does not have the option of masking parts of the CCD for shutterless operation.
Each set of observations consists of 15 frames, each of which was acquired using eight sub-exposures of $0.05~\mathrm{s}$, leading to a total effective exposure time of $0.41~\mathrm{s}$ per frame.
Because of a delay incurred by the CCD readout, it takes approximately $10~\mathrm{s}$ to complete the acquisition of the Stokes vector.
The observations are binned $2\times2$ on-chip to increase S/N per pixel.
We performed simultaneous scans of the same field of view (FOV) with the spectropolarimeter (SP) in ``fast map'' mode (with limited spatial resolution, but slightly increased sensitivity), to measure the strong, signed magnetic fields, using the Zeeman effect in the \FeI\ line pair at $630~\mathrm{nm}$.

Data were obtained for various positions along the solar limb: the north (N) limb, the north-west (NW) limb, and the east (E) limb.
These complementary data sets are used to verify the measurement of a solar scattering polarization signal (and not an instrumental effect), and to assess systematic effects in the linear Stokes parameters ($U/I$, $Q/I$) that should not contain scattering polarization.
Henceforth we employ the global Stokes coordinate system of \textit{Hinode} that has $+Q$ oriented parallel to the N limb.

All raw frames are run through the standard \textit{Hinode} pipeline\footnote{The \textit{Hinode} pipeline is implemented in the IDL SSW, see \href{http://www.lmsal.com/solarsoft/}{\texttt{http://www.lmsal.com/solarsoft/}}.} that performs dark subtraction, flat fielding, cosmic ray hit removal, and polarimetric demodulation.

\subsection{\NaID\ observations}
During the same observing sequence, full Stokes line scans of the \NaID\ line were also performed with the NFI instrument.
\NaID\ is known to have a scattering polarization of $\sim0.1\%$ \citep{sss} with an antisymmetric signature in the line wings (due to interaction with the D$_2$ line), and often with a peaked line core polarization of yet unknown origin \citep{StenfloNa, D1core, Casinidoesd1}.
Filter observations in the D$_1$ and D$_2$ lines with limited spatial resolution were performed by \citet{Stenflospatial}, without measuring significant scattering polarization signatures in the D$_1$ line.

Unfortunately, the \textit{Hinode} line scan observations in the D$_1$ line exhibit instrumental effects at the $10^{-3}$ level, which prevents us from detecting real scattering polarization signals. 
A polarization that varies with $\mu$ is detected in $Q/I$ at the N limb, but it does not rotate in polarization direction at the other limb positions, indicating that it is an instrumental effect that causes a print-through of the intensity limb darkening.
In all Stokes $U/I$ observations, a blotchy, varying instrumental polarization pattern is measured.
The origin of these two effects is currently unknown.


\begin{figure*}[!ph]
	\includegraphics[]{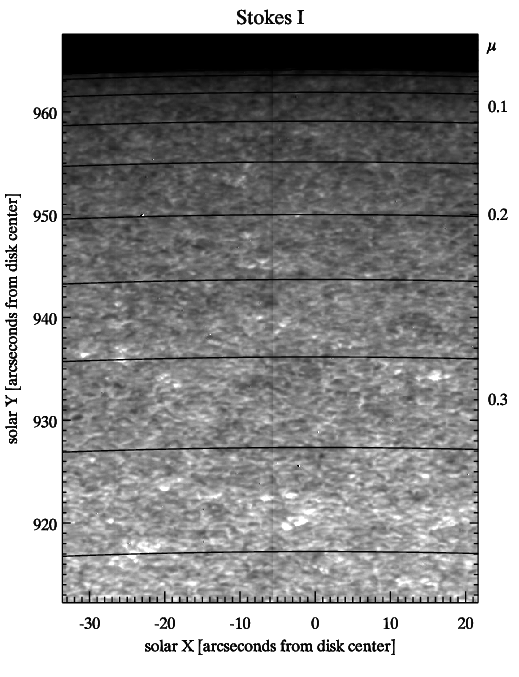}
	\hskip10mm
	\includegraphics[]{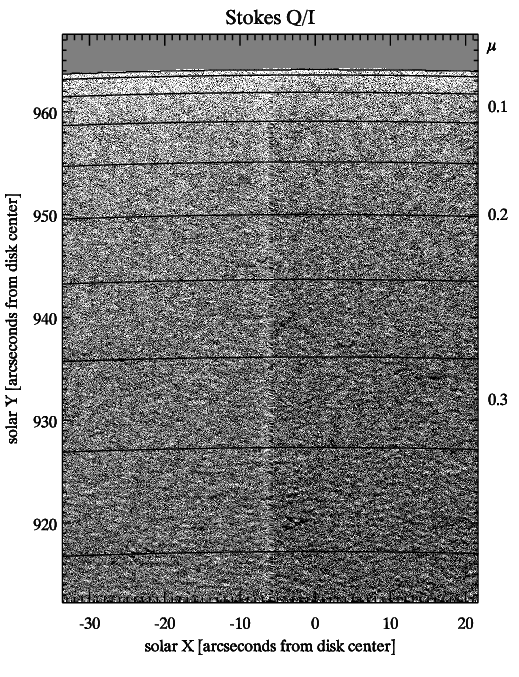}\\
	\includegraphics[]{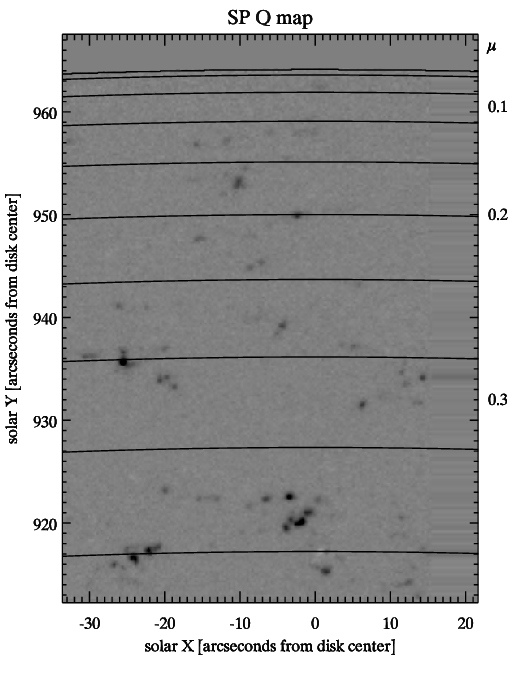}
	\hskip10mm
	\includegraphics[]{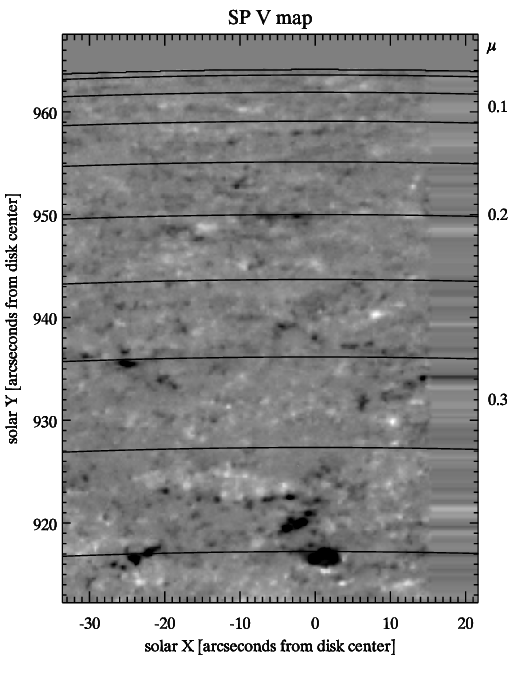}
	\caption{Stokes $I$ and $Q/I$ observations in the CN band at 388 nm at the north limb of the Sun. The bin limits of $\mu$ used in the data reduction are overplotted as contours. A single intensity frame is presented, which exhibits clearly resolved granules down to $\mu=0.2$, and a granular pattern down to $\mu=0.1$. The $Q/I$ data is the average of all 15 acquired frames and is clipped between $-0.3$ and $0.3\%$. A clear, positive scattering polarization signal is observed. Spurious polarization signals due to residual image motion are present in the data at positions with a steep intensity gradient. The conspicuous vertical line in the center of the CN band images is due to the fact that the detector consists of two monolithic parts that meet in the middle of the FOV. The pixels surrounding this line are masked during data reduction. The lower two panels present the data from the SP at 630 nm, which show the large-scale magnetic structures, mostly associated with polar faculae. The Stokes $Q$ map mostly shows vertical magnetic fields, whereas the Stokes $V$ map shows more horizontal magnetic fields of mixed polarity.}
	\label{CNdata}
\end{figure*}

\section{Data reduction}\label{datareduction}

\subsection{Instrumental polarization}
As the BFI instrument has never been polarimetrically calibrated, we extrapolated the polarimetric measurement model of \citet{Ichimotocal} to the observing wavelength of $388~\mathrm{nm}$.
The retardance of the rotating wave plate is known at this wavelength to be 9.338 waves.
All measured Stokes parameters have an opposite sign in the BFI data compared to the NFI, because the polarizing beam-splitter analyzes linear polarization with a direction at $90^\circ$ with respect to the NFI.

The theoretical $\tens{X}$ matrix \citep[see][]{Ichimotocal} for the shuttered CN band observations, which describes the polarimetric efficiencies with its diagonal elements and cross-talk with its non-diagonal elements and takes into account the exposure intervals during the continuous modulation, is calculated to be
\begin{equation}
	\tens{X}_\mathrm{388}=\left(\begin{array}{rrrr}
		\phantom{-}1.0000 & -0.1956 & \phantom{-}0.0000 & \phantom{-}0.0000\\
		0.0000 & -0.4323 & 0.0000 & 0.0000\\
		0.0000 & 0.0000 & 0.4323 & 0.0000\\
		0.0000 & 0.0000 & 0.0000 & 0.4549
	\end{array}\right)\,.
	\label{Xmatrix}
\end{equation}
The $Q\rightarrow Q$ and $U\rightarrow U$ elements indicate that linear polarization is modulated with reasonable efficiency. 
The only non-zero non-diagonal element is $Q \rightarrow I$, which is caused by the polarizer in the $-Q$ direction in combination with the fact that the modulator is not a half-wave plate at this wavelength.
The data were corrected for this $Q \rightarrow I$ cross-talk, but the effect of it is minor because $Q \ll I$.
The accuracy of this matrix is quoted by \citet{Ichimotocal} to be $\sim$0.05 for the diagonal elements, and $\sim$$10^{-3}$ for the non-diagonal elements, although specific instrumental polarization problems at this wavelength cannot be directly assessed.
The efficiency of the polarization measurements may also be reduced because the polarizing beam-splitters are probably not perfectly polarizing at this wavelength.
This made us in the end quantify our results with a relative polarization measure, which is independent of the polarimetric efficiency.
Instrumental polarization $I\rightarrow Q$ and $I\rightarrow U$ of $\sim$$10^{-3}$ was measured in the CN band data sets by applying curve fits with variable offsets to the scattering polarization data.
This instrumental polarization is small enough so that second-order effects that create a print-through of the intensity signal in $Q/I$ and $U/I$ \citep[see][]{Keller96} are very small and the last term of Eq.~\ref{effects} can be safely neglected.

The CN band data at the N limb after correction of instrumental polarization (cf.~Eq.~\ref{Xmatrix}) are presented in Fig.~\ref{CNdata}.
A clear scattering polarization signal is detected parallel to the limb, which increases towards the limb.
Similar signals are also detected at the NW and E limb, with the correct direction of linear polarization (i.e., $-U$ and $-Q$, respectively), confirming the solar origin of the signal. 
The pointing stability at the E limb proved to be poor, so we discarded that data set.

We chose not to rotate the local Stokes coordinate system so that $Q/I$ would always be oriented parallel to the limb to prevent the introduction or mixing of systematic effects.
All measurements are therefore presented in the global $[Q,U]$ coordinate system of \textit{Hinode} \citep[see][which incidentally has the opposite sign of Stokes $U$ as compared with the current \textit{Hinode} pipeline]{Ichimotocal}.

\subsection{Masking}
Various data masks were created to enable a statistical analysis of the N and NW data sets.
First, the pixels at the edges of the two CCD halves were flagged and discarded, as were dead pixels.
The obtained SP scans were converted to line-of-sight and plane-of-sky total line polarization maps with the \texttt{stksimages} routine included in the \textit{Hinode}-SP pipeline.
Then these maps were warped to the BFI coordinate system and sampling with the equations of \citet{Centenowarp}.
The SP results for the N limb are presented in the lower panels of Fig.~\ref{CNdata}.
Because we are primarily interested in the polarization signals in the internetwork quiet Sun, all pixels exceeding $3\sigma$ of the noise in the transverse magnetograms were excluded.
The spatial resolution of the fast map SP data is course enough to accommodate co-registration errors between the SP data and the BFI data.
Furthermore, it is expected that formation height differences between the CN band and in \FeI\ $630.15$ and $630.25~\mathrm{nm}$ make features appear at slightly different locations (shifted in the limbward direction) when observing close to the limb, where the atmospheric stratification is observed almost perpendicularly.
The spatial resolution of the SP data ($0.32 \arcsec$) is courser than these expected shifts ($\sim 0.1 \arcsec$).
To a large extent these magnetic masks only take out the polar faculae that are also apparent in the CN band intensity data of Fig.~\ref{CNdata}.
Because the intensity of these faculae is much higher than the surroundings, the magnetic masks also serve to decrease systematic cross-talk effects from Stokes $I$ to $Q/I$ or $U/I$.
The remaining pixels are now assumed to represent the quiet Sun.
Polarization signals are observed practically everywhere in the longitudinal SP maps \cite[see also][]{Tsuneta}.
Because of the limited spatial resolution of the fast map SP data, magnetic structure at granular scales cannot be distinguished very close to the limb.

We assigned a value for $\mu\equiv\cos\theta$ to each pixel.
These $\mu$ masks are used to sort the data in bins of $\mu$ intervals to produce plots of the CLV of the CN band scattering polarization as a function of $\mu$.
We chose a bin size for $\mu$ of $0.04$, which is a compromise between sufficient sampling of the CLV and obtaining sufficient statistics on granules and pixel noise within each bin.
The bin limits are overplotted as contours on the N limb data of Fig.~\ref{CNdata}.
Because the scattering polarization decreases rapidly with increasing limb distance, but the bins grow in size, the relative noise of the measured scattering polarization does not vary rapidly.
After binning, we integrated over all pixels in a $\mu$ bin to increase the S/N of the measured scattering polarization.
In this way, specific spatial information is lost, except for a dedicated averaging over `granules' and `intergranules', as explained below.

We fit the limb darkening in the quiet sun data as a function of $\mu$ (without binning) with a 5th-degree polynomial $\langle I\rangle(\mu)$ \cite[cf.][]{Neckel}. 
The `granules' and `intergranules' are subsequently selected to have a higher and lower intensity than the median value of the limb-darkening corrected intensity data, respectively. 
Any definition of `granules' and `intergranules' is necessarily somewhat arbitrary.
The median value is chosen to balance the number statistics between `granules' and `intergranules', and also to deal with spurious polarization signals, as elaborated in the next subsection.

The granular pattern is clearly resolved down to $\mu=0.2$ as shown in Fig.~\ref{CNdata}.
The granulation close to the limb appears foreshortened.
The bright parts are in fact the hot granular walls \citep[see][]{KellerMHD}.
This means that the identification of `granules' and `intergranules' does not necessarily coincide with a similar identification at disk center.
The terms `granules' and `intergranules' used throughout this paper should therefore be interpreted with this caveat in mind.

\subsection{Spurious polarization signals}
Spatially varying, spurious polarization signals are clearly observed in the $Q/I$, $U/I$ and also the $V/I$ data, similar to the structure in the top right panel of Fig.~\ref{CNdata}. 
A large fraction of these signals is caused by cross-talk between the Stokes parameters introduced through image motion during the modulation, which is not completely corrected by the image stabilization system of the \textit{Hinode}-SOT.
Note that at the high polarimetric sensitivity levels of these observations, a minute displacement of 1~milliarcsecond already creates a detectable signature in the polarization data. 

We assume that the displacements are indeed small and constant over the FOV, and modeled this spurious signal due to residual image motion similar to \citet{lites87}.
Let $\tens{M}$ and $\tens{D}$ denote the modulation matrix and the demodulation matrix.
In a single-beam instrument, cross-talk from $I$ will dominate.
For $Q$, we then have
\begin{equation}
	\Delta Q = \sum_{k}\left(\tens{X}^{-1}\cdot\tens{D}\right)_{2,k}\,\tens{M}_{k,1}\,(\nabla I\cdot\vec{r}_k)\,,
\end{equation}
where $\vec{r}_k$ is the image displacement at stage $k$ of the modulation cycle.
Because $\tens{X}$, $\tens{D}$, and $\tens{M}$ are constant, and $\nabla I$ is constant to good approximation during the modulation cycle, we can write
\begin{equation}
	\Delta Q = \nabla I\cdot\sum_{k}c_{2,k}\,\vec{r}_k = \nabla I\cdot\vec{r}_Q\,,
	\label{gradient}
\end{equation}
where $\vec{r}_Q$ is a vector in the coordinate system of the image that is proportional to the overall image motion during modulation as averaged and weighted by the modulation sequence.
The derivations are analogous for Stokes $U$ (and $V$), but for now we describe the scattering polarization at the N limb, which is detectable in Stokes $Q$.

We remove effects of image motion during the modulation cycle using an $\vec{r}_Q$ found for each frame by minimizing $\langle(Q_\mathrm{meas.}-\Delta Q)^2\rangle$ in an area away from the limb, where the angled brackets denote averaging over that area.
Note that with Eq.~\ref{effects}
\begin{equation}
	(Q_\mathrm{meas.}-\Delta Q)^2=Q'^2 + (Q_\mathrm{i.m.} - \Delta Q)^2 + 2\,Q'\,(Q_\mathrm{i.m.} - \Delta Q)\,,
\end{equation}
where $Q' = Q_\mathrm{real}+Q_\mathrm{s.e.}+\tens{X}_{21}\,I$.
In areas where limb darkening is small, $\Delta Q$ vanishes after averaging and the last term is constant.
Residual spurious polarization patterns are still present, probably due to differences between the numerical derivatives of the intensity data and the real gradients.

In the absence of limb darkening, the average value of $Q$ in the `granules' and the `intergranules' is not affected by a correction according to Eq.~\ref{gradient}, but the spread of the distribution around this average is decreased. 
This is because the definition of `granules' and `intergranules' as pixels exhibiting more and less intensity than the median is such that after adding up all selected pixels, the residual value of $Q_\mathrm{i.m.}$ is zero: for both the `granules' and the `intergranules' mask it constitutes an integration of the derivative (gradient) of the intensity image over an interval with identical intensity values at the boundaries. 

The influence of the limb darkening is calculated as
\begin{equation}
	\left\langle Q_\mathrm{i.m.}\right\rangle_\mathrm{g,i}(\mu)= \frac{1}{N_\mathrm{g,i}} \iint_\mathrm{g,i} \vec{r}_Q\cdot \nabla \left(I_\mathrm{rel.}(x,y) + \langle I\rangle(\mu)\right)\,\mathrm{d}x\,\mathrm{d}y,
\end{equation}
which is a normalized integration over all `granules' (g) or `intergranules' (i) over the bin with average value $\mu$, containing $N_\mathrm{g}=N_\mathrm{i}$ pixels.
$I_\mathrm{rel.}$ is the granulation intensity pattern after limb darkening correction $\langle I\rangle(\mu)$. 
As explained above, the first term of the integral disappears and the equation reduces to
\begin{equation}
	\left\langle Q_\mathrm{i.m.}\right\rangle_\mathrm{g,i}(\mu)\approx \frac{f_\mathrm{g,i}}{N_\mathrm{g,i}/\Delta x}\hat{r}_y\,\left(\langle I\rangle(\mu_2)-\langle I \rangle(\mu_1)\right)\,,
\end{equation}
where $[\mu_1,\mu_2]$ is an interval around $\mu$.
To account for the geometry of the `granules' and `intergranules' in the bin, we introduced a ``filling parameter'' $f_\mathrm{g,i}$. 
Only the component $\hat{r}_y$ of the (random) image motion $\vec{r}_Q$ in the direction of decreasing $\mu$ (in this case increasing $y$) contributes to a residual signal after integration.
The integration over the perpendicular spatial coordinate $x$ only results in a factor $\Delta x$ (the number of pixels in the $x$ direction) to the normalization with the amount of selected pixels $N_\mathrm{g,i}$.
As a result of our definition, `granules' and `intergranules' are spatially interspersed with identical number densities, so that we have $f_\mathrm{g}=f_\mathrm{i}=0.5$ and hence $\langle Q_\mathrm{i.m.}\rangle_\mathrm{g}(\mu)=\langle Q_\mathrm{i.m.}\rangle_\mathrm{i}(\mu)$.
Also, because the residual image motion $\vec{r}_Q$ (and therefore $\hat{r}_y$) is most likely random in magnitude and orientation, this leads to further averaging out of the influence of image motion on the observed difference in measured polarization.
Therefore we concluded that the spurious polarization due to image motion and its correction procedure according to Eq.~\ref{gradient} \emph{on average} does not introduce artificial polarization differences between `granules' and `intergranules'.

The remaining spurious signals are due to solar evolution and are difficult to model.
They are, however, very relevant for our investigation: during the polarimetric modulation ($\sim$10 s) all pixels change intensity because of the dynamic granulation.
Therefore this effect may create an artificial difference in $Q/I$ and $U/I$ for `granules' and `intergranules'. 
This difference is expected to be very similar in magnitude for Stokes $Q$ and Stokes $U$, because they are \mbox{(de-)modulated} with identical amplitude.
However, after correction for the modulation phase delay the demodulation sequence is slightly different for $Q$ and $U$.
The center-to-limb behavior of these spurious signals for `granules' and `intergranules' is investigated for $U/I$ in the N limb data set, where ideally only scattering polarization signals are present in $Q/I$ and vice versa with the data at the NW limb.
With the slow variation of the appearance of the granulation away from disk center, the difference between $\langle Q_\mathrm{s.e.}\rangle_\mathrm{g}(\mu)$ and $\langle Q_\mathrm{s.e.}\rangle_\mathrm{i}(\mu)$ is expected to be a smooth function.

The real solar data can now be referenced to the instrumental polarization and spurious signals due to solar evolution (and other unknown effects). 
With the assumptions and procedures described in this subsection, Eq.~\ref{effects} is converted to
\begin{equation}
	\Delta_\mathrm{g,i}\langle Q_\mathrm{real}\rangle(\mu)=\Delta_\mathrm{g,i}\langle Q_\mathrm{meas.}\rangle(\mu)-\Delta_\mathrm{g,i}\langle Q_\mathrm{s.e.}\rangle(\mu)-\tens{X}_{21}\,\Delta_\mathrm{g,i}\langle I\rangle(\mu)\,,
	\label{reduction1}
\end{equation}
where $\Delta_\mathrm{g,i}$ is used to denote the difference between `granules' and `intergranules'.
The difference in polarization degree for average `granules' and `intergranules' is f determined or each bin through
\begin{equation}
	\Delta_\mathrm{g,i}\left\langle\frac{Q_\mathrm{real}}{I}\right\rangle(\mu)=\Delta_\mathrm{g,i}\left\langle\frac{Q_\mathrm{meas.}}{I}\right\rangle(\mu)-\Delta_\mathrm{g,i}\left\langle\frac{Q_\mathrm{s.e.}}{I}\right\rangle(\mu)\,,
	\label{reduction2}
\end{equation}
assuming that $I_\mathrm{real}=I_\mathrm{meas.}=I$.
The instrumental polarization (the last term in Eq.~\ref{reduction1}) does not introduce differences in polarization degree, as it only adds a flat background in $Q/I$.
However, to determine a \emph{relative} difference between the scattering polarization of `granules' and `intergranules', it needs to be properly taken into account.


\section{Results and discussion}\label{results}

\begin{figure*}[t]
	\includegraphics{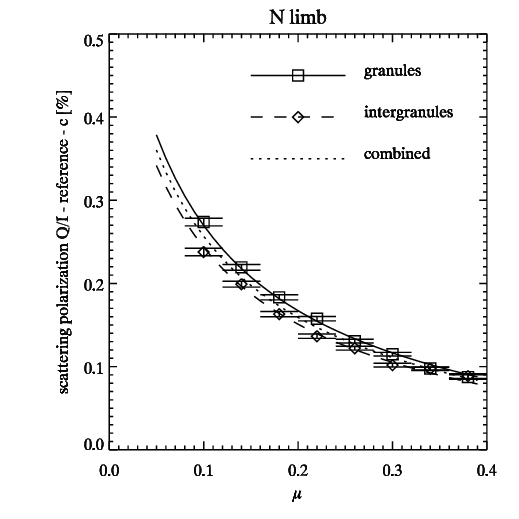}
	\hskip10mm
	\includegraphics{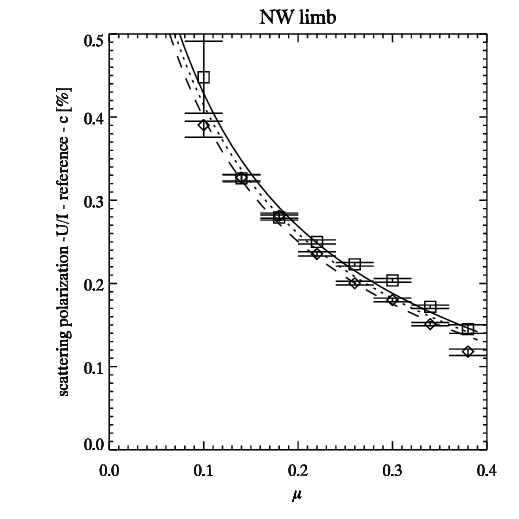}\\
	\includegraphics{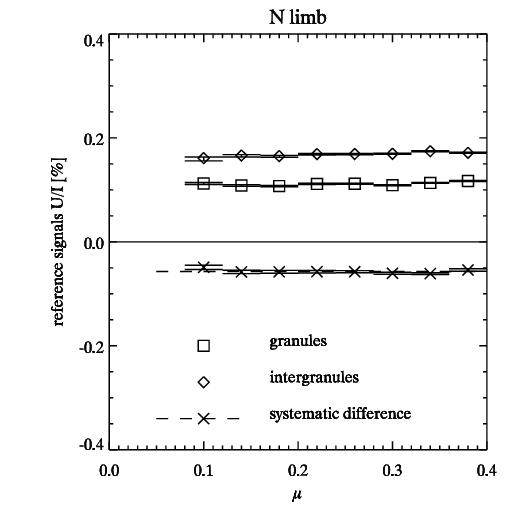}
	\hskip10mm
	\includegraphics{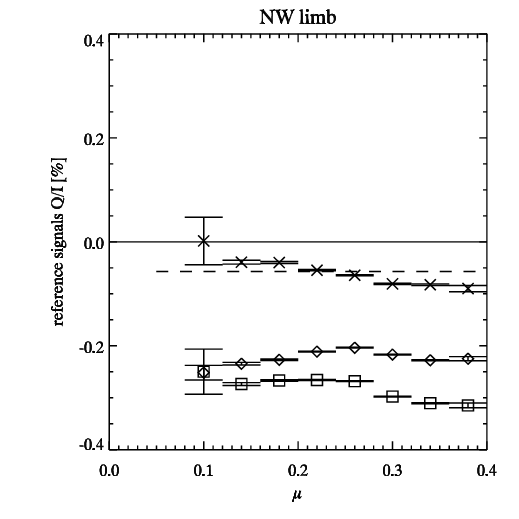}\\
	\caption{Results for the center-to-limb variation of the CN band scattering polarization data at the N and NW limb. The upper panels present the measured scattering polarization after splitting the binned data according to `granules' and `intergranules' and subtracting the obtained difference between `granules' and `intergranules' in the reference data (shown in the lower panels) and the polarization offset $c$. The CLV fit of the combined data of the scattering polarization is represented by the dotted curve. The results for `granules' and `intergranules' are represented separately by the solid and dashed curves, respectively.}
	\label{resultsCLV}
\end{figure*}

The $Q/I$ and $U/I$ data as a function of $\mu$ for the N and NW limb are presented in Fig.~\ref{resultsCLV}.
The observed CLV of the scattering polarization is plotted in the upper panels, after splitting the data according to `granules' and `intergranules', and referencing it to the observed difference in the $U/I$ data at the N limb.
The reference signals for the scattering polarization signals are plotted in the lower panels of Fig.~\ref{resultsCLV} and are defined as the linear polarization measurements for the (normalized) Stokes parameter that should not contain scattering polarization, so $U/I$ for the N limb data and $Q/I$ for the NW limb data.
The absolute values for the reference signals are irrelevant, because the polarization offset $I \rightarrow Q,U$ is poorly known at this level for both $Q/I$ and $U/I$.
The reference signals exhibit a difference of similar magnitude for the N and NW data.
Yet, the systematic difference is flat with $\mu$ at the N limb, whereas it slightly varies with $\mu$ at the NW limb.
The reason for this may be somehow related to the fact that the rectangular FOV is not aligned with the NW limb direction, whereas at the N limb the FOV is oriented much more symmetrically with respect to the scattering polarization signals.
Because the magnitude of the reference signals is the same, and because $Q$ and $U$ are modulated with identical amplitude, we assumed that the difference in the reference signals (e.g.,~due to solar evolution) is the same for both data sets.
The reference signals at the N limb (in the lower left panel of Fig.\ref{resultsCLV}) are obviously the cleanest, and we adopted the average difference measured in $U/I$ as a reference for the scattering polarization data for `granules' and `intergranules'.

The error bars in the plots of Fig.~\ref{resultsCLV} are the $1\sigma$ ``standard errors'' of the data within each bin.
The standard error is defined as
\begin{equation}
	\sigma=\frac{s}{\sqrt{N}}\,,
\end{equation}
where $s$ is the sample standard deviation and $N$ is the number of samples, and represents the accuracy with which the average value in each bin is known, provided the noise is Gaussian.
Because systematic effects are still present in the bins (e.g., variation of polarization within the bins, residual spurious signals), these error bars merely serve as an indication of the uncertainty in the averaged data points.
Therefore the error bars could not be used to quantify the significance of the differences between scattering polarization in `granules' and `intergranules', and we needed to resort to combining data at different $\mu$ values.
The deviations of the averaged data-points as a function of $\mu$ then need to be critically assessed whether they appear random or systematic.

After subtracting the found average polarization difference found for the reference signals, the CLV of the scattering polarization in both data sets appears to be smooth (upper panels of Fig.~\ref{resultsCLV}).
Moreover, apparent differences between the CLV of `granules' and `intergranules' are readily observed in both cases.
In general, the difference between the scattering polarization signals appear to scale with the total scattering polarization, with the signals in the `granules' being larger.
This indicates that the origin of this difference is likely to be solar, because many instrumental effects and solar evolution introduce an absolute difference, whereas physical effects on the Sun, such as differences in scattering anisotropy and Hanle depolarization, introduce mostly relative effects.

Note that the absolute value of the measured scattering polarization is about 40\% higher at the NW compared to the N limb.
The cause for this is either a reduced value of the $Q \rightarrow Q$ in the $\tens{X}$ matrix, or a real solar effect such as overall depolarization due to stronger turbulent magnetic fields at the N pole.
The only remarkable difference in the corresponding SP maps is the presence of polar faculae at the N limb \cite[see also][]{Tsuneta}.
Incidentally, such a systematic variation along the limb was not found in observations of scattering polarization in C$_2$ lines by \citet{Kleintsynoptic}.
In any case, this difference in absolute value once more makes it necessary to characterize the difference in scattering polarization between `granules' and `intergranules'  as a relative difference, i.e.,
\begin{equation}
	\Delta_\mathrm{g,i, rel.}\left\langle\frac{Q_\mathrm{real}}{I}\right\rangle=\frac{\Delta_\mathrm{g,i}\left\langle\frac{Q_\mathrm{real}}{I}\right\rangle}{\left\langle\frac{Q_\mathrm{real}}{I}\right\rangle}\,.
	\label{eq_reldif}
\end{equation}

To obtain a first quantitative measure of the differences in scattering polarization between `granules' and `intergranules', and, more importantly, establish the zero point of the polarization scale, we performed curve fits of the CLV of the measured scattering polarization according to \citet{StenfloCLV},
\begin{equation}
	\frac{Q}{I},\frac{U}{I}=\frac{a\,(1-\mu^2)}{\mu-b}+c\,,
\end{equation}
in which $a$ represents the amplitude of the scattering polarization, and $c$ fits the unknown offset of the polarization scale, which is the main goal of this fitting procedure.
The fit parameter $b$ was introduced by \citet{StenfloCLV} to improve the fits of the CLV of many lines exhibiting scattering polarization.
First, the combined data was fitted.
The results of these fits are shown in the top panels of Fig.~\ref{resultsCLV} with dotted curves.
Note that inclusion of the limb darkening function in the fit, as suggested by \citet{StenfloCLV} and references therein, did not improve the fit quality and in fact slightly increased the 1$\sigma$ errors on the determined parameters $c$.
Then the data for `granules' and `intergranules' were fit separately with $b$ and $c$ obtained from the previous fit as constants.
The results are also shown in the upper panels of Fig.~\ref{resultsCLV} as the solid and dashed curves, respectively.
The fit yields $a_g=0.048\pm0.006$ and $a_i=0.044\pm0.006$ for the N limb data, and $a_g=0.079\pm0.007$ and $a_i=0.074\pm0.006$ for the NW data.
After propagation of the ``standard errors'' according to a Gaussian treatment (assuming that the errors are uncorrelated), the relative differences between the amplitudes of the scattering polarization fits of `intergranules' and `granules' as defined in Eq.~\ref{eq_reldif} are found to be $(-9.7\pm16.8)\%$ for the N limb, and $(-6.8\pm11.3)\%$ for the NW limb.
For clarity: the minus sign indicates a lower scattering polarization signal in the `intergranules'.
The larger relative error at the N limb is due to the lower absolute measured scattering polarization signal, which renders the curve fits more sensitive to the unknown polarization offset.
The obtained $1\sigma$ errors are not reliable, because the original errors are likely dominated by systematic errors, which are not taken into account.
Moreover, the fitting model is quite simplistic and inherently assumes that a single scattering behavior is present for the masked `granules' and `intergranules' pattern as a function of $\mu$, whereas the observed granulation pattern clearly changes appearance.
Also, if the Hanle effect has a significant impact on the scattering polarization, systematic effects are created by large-scale spatially varying magnetic fields.
These fit results therefore comprise at best a first estimate of the relative difference.
To further constrain the measured differences, we have to analyze the data of the `granules' and `intergranules'  more localized in $\mu$.

\begin{figure}[t]
	\includegraphics[]{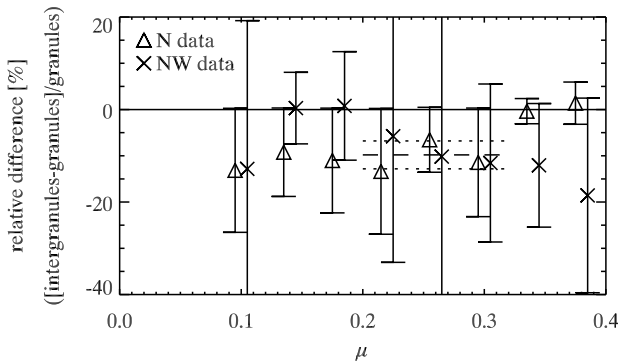}
	\caption{Relative differences between `granules' and `intergranules' for the data from Fig.~\ref{resultsCLV}. The data were shifted by $\mu=\pm0.05$ to avoid overlapping of error bars. The error bars are the propagated ``standard errors'' and do not take systematic effects into account. The dashed line represents the average measure relative difference for $0.2<\mu\le0.3$ at both data sets, with the dashed lines as the $1\sigma$ range.}
	\label{reldif}
\end{figure}

The relative differences between `granules' and `intergranules' as a function of $\mu$ for both data sets is plotted in Fig.~\ref{reldif}.
The error bars are the propagated ``standard errors'', which are treated as Gaussians.
So again, these error bars do not properly incorporate systematic effects, but give a rough indication of the total error of the data points.
It is clear from this plot that the average of most (if not all) measured relative differences are $\le0$, indicating a significant decrease of scattering polarization in the `intergranules'.
It is however also obvious that these average values exhibit systematic effects as a function of $\mu$.
To analyze granulation that is clearly resolved by the observations, we now discarded the data for $\mu<0.2$.
Furthermore, the N limb data above $\mu=0.3$ is conspicuously different from the rest of the data as the difference in scattering polarization between `granules' and `intergranules' between $\mu=0.3$ and $\mu=0.4$ is fully consistent with zero, whereas all other data points (including those between $\mu=0.3$ and $\mu=0.4$ at the NW limb) are distributed fairly smoothly around a value below zero.
We therefore computed the average value for both data sets between $\mu=0.2$ and $\mu=0.3$, where the data in Fig.~\ref{reldif} appears to behave in a relatively random fashion, and, moreover, is consistent between the N and NW data sets.
Also, the reference signals in the lower panels of Fig.~\ref{resultsCLV} are most consistent in this region.
The obtained average plus/minus the standard deviation thereof are overplotted as a dashed and dotted lines, respectively.
The obtained value for the relative difference of scattering polarization between `intergranules' and `granules' for $0.2<\mu\le0.3$ is $(-9.8\pm3.0)\%$.
Of course, systematic effects cannot be completely excluded either for this result which used a limited amount of data, but we adopted it as our main result that is least affected by systematic effects.

Note that also most data outside the chosen $\mu$ range match the adopted relative difference, except for the four data-points that correspond to zero difference.
Interestingly, the two ``zero'' data points at the N limb correspond to a region of enhanced facular activity, which also harbors strong magnetic fields outside the apparent faculae according to the longitudinal (Stokes $V$) SP map of Fig.~\ref{CNdata}. 
The observations at the NW limb at those values of $\mu$ harbor much weaker magnetic fields.
If the Hanle effect is the dominating factor for the absolute line polarization, then this implies that at this region close to polar faculae, both the granules and the intergranules are permeated by strong magnetic fields.
Because of its proximity to the limb, it is difficult to assess from the SP data whether there is increased magnetic activity at the values of $\mu$ at the NW limb that exhibit zero difference in scattering polarization.
But it is obvious from the data in Fig.~\ref{resultsCLV} that the absolute values of the scattering polarization are also consistently lower there.

It is worth remarking that the observed absolute difference between `granules' and `intergranules' of $\sim$$0.04\%$ in Fig.~\ref{resultsCLV} corresponds well with a Hanle depolarization difference from $11~\mathrm{G}$ to $>25~\mathrm{G}$ in Fig.~8 of \citet{Shapiro}, which matches the prediction by \citet{TBnature}.
It is, however, required to compare the data presented herein to an MHD model with realistic line synthesis of the CN band before any final statements about the nature of the weak, turbulent magnetic fields can be made.
The results presented in Figs.~\ref{resultsCLV} and \ref{reldif} are therefore most useful as constraints to future models.


\section{Conclusions}\label{conclusions}
We conclude that there is a significant difference in the scattering polarization signals between `granules' and `intergranules' (as defined in Sect.~\ref{datareduction}) in the CN band as shown in Fig.~\ref{reldif}.
It is determined that the `intergranules' exhibit $(9.8\pm3.0)\%$ less scattering polarization for $0.2<\mu\le0.3$.
The physical cause for this difference (formation effects, scattering anisotropy variations, different turbulent magnetic field strengths) cannot be established from these observations, but could be obtained by confronting the obtained data with a realistic MHD model combined with detailed polarized line synthesis of the CN band.

Two polarization effects are observed in the data that correlate with the presence of polar faculae: the overall scattering polarization at the N limb is lower than at the NW limb, and the difference in scattering polarization between `granules' and `intergranules' at a region of strong facular activity is zero.
Both these observations could be understood in terms of the Hanle effect, indicating the presence of stronger (turbulent) magnetic fields near the polar faculae.
Unfortunately, the CN band and spectropolarimeter observations are dominated by noise and are moreover not strictly cotemporal.
This seriously complicates the interpretation of correlation analyses between both data sets.
Therefore no conclusion can be drawn as to whether or not the magnetic fields measured by the SP have any significant effect on the CN band scattering polarization. 

It is impossible to completely exclude systematic effects and assess their influence on the obtained results.
One way of addressing the systematic effects is to construct a complete physical model for the reference signals in the lower panels of Fig.~\ref{resultsCLV}.
But more trust in the data can also be obtained by showing that the possible physical effects on the Sun indeed have a similar scaling effect on the scattering polarization, as observed in the data after correcting with the (additive) reference signals.

On the one hand, the CN band is well suited for scattering polarization observations because its absolute scattering polarization signals are relatively large, mostly due to the rapid increase of the continuum polarization towards the blue \citep[see][]{TBcont}.
On the other hand, the radiative transfer in the (molecular) CN band is relatively difficult to model compared to some other atomic lines.
More suitable for observations of scattering polarization at high spatial resolution are therefore lines like \mbox{Ba~\sc{ii}} at $455.4~\mathrm{nm}$, \mbox{Sr~\sc{i}} at $460.7~\mathrm{nm}$, and \mbox{Na~\sc{i}~D$_2$}, all of which have a sizable scattering polarization ($\sim1\%$) and are relatively well understood.
A narrow-band tunable filter needs to be employed for these narrow lines. 
This introduces the possibility to distinguish line and continuum polarization, which is not possible for the CN band observations.
Because no such capabilities are planned in space in the foreseeable future, observations of scattering polarization at high spatial resolution need to be attempted from the ground to further constrain models of weak, turbulent magnetic fields in the solar photosphere.


\begin{acknowledgements}
We thank Roberto Casini for useful discussions.
Hinode is a Japanese mission developed and launched by ISAS/JAXA, collaborating with NAOJ as a domestic partner, NASA and STFC (UK) as international partners. Scientific operation of the Hinode mission is conducted by the Hinode science team organized at ISAS/JAXA. This team mainly consists of scientists from institutes in the partner countries. Support for the post-launch operation is provided by JAXA and NAOJ (Japan), STFC (U.K.), NASA, ESA, and NSC (Norway). 
\end{acknowledgements}

\end{document}